\documentclass[twocolumn,showpacs,preprintnumbers,amsmath,amssymb]{revtex4}
\usepackage{graphicx}
\usepackage{dcolumn}
\usepackage{bm}
\begin{document}
\title{Interplay between Zeeman interaction and spin-orbit coupling in a
two-dimensional semiconductor system}
\author{Manuel Val\'{\i}n-Rodr\'{\i}guez}
\affiliation{
Departament de F\'{\i}sica, Universitat de les Illes Balears,
E-07122 Palma de Mallorca, Spain}
\author{Rashid G.~Nazmitdinov}
\affiliation{Departament de F{\'\i}sica,
Universitat de les Illes Balears, E-07122 Palma de Mallorca, Spain}
\affiliation{Max-Planck-Institut f\"{u}r Physik  komplexer Systeme,
D-01187, Dresden, Germany}
\affiliation{Bogoliubov Laboratory of Theoretical Physics,
Joint Institute for Nuclear Research, 141980 Dubna, Russia}
\date{\today}

\begin{abstract}
We analyse the interplay between Dresselhaus, Bychkov-Rashba, 
and Zeeman interactions in a two-dimensional semiconductor quantum system
under the action of a magnetic field. When a vertical magnetic field is 
considered, we predict that the interplay results in an effective cyclotron 
frequency that depends on a spin-dependent contribution. For in-plane 
magnetic fields, we found that the interplay induces an anisotropic
effective gyromagnetic factor that depends on the orientation of the applied field
as well as on the orientation of the electron momentum.
\end{abstract}


\maketitle

The effects produced by different spin-dependent interactions in semiconductor
structures are currently in the forefront of experimental and theoretical efforts
in mesoscopic physics. The  explosive activity is motivated by the desire for a deep
understanding of quantum coherence phenomena. The other driving force is the hope 
that the spintronics research would provide novel, low-dissipative microelectronic
devices \cite{wolf, fab}.

Most  measurements of spin effects in semiconductor microstructures are performed 
applying a magnetic field.  In this case, two intrinsic spin-dependent interactions 
naturally appear: the well-known Zeeman interaction, which couples the electron 
spin with the applied field, and the spin-orbit coupling. The latter has 
been extensively studied starting from two-dimensional electron gas  to quantum 
dots \cite{winbo,deste,naz,chen,zul,vos}, and constitutes the basis of several 
proposals for spin-based applications \cite{datta, ohe, schlie}.

In a semiconductor,  the spin-orbit coupling originates as 
a relativistic effect caused by electric fields present in the material
(cf \cite{Ra}). Such a system can be described by the Hamiltonian
\begin{equation}
\label{ham}
{\cal H}=T
+{\cal H}_D+{\cal H}_R+{\cal H}_Z
\end{equation}
Here $T=\left(p_x^2+p_y^2\right)/2m^*$ is the kinetic energy with the effective
mass $m^*$ of electrons in the conduction band of the sample and $p_x,p_y$ are 
components of the momentum. The  electric field caused by the
bulk inversion asymmetry of the crystalline structure contributes to the
Hamiltonian, Eq.(\ref{ham}), with the 
Dresselhaus term,  ${\cal H}_D$ \cite{dress}. 
This term has, in general, a cubic dependence on the momentum of the carriers. 
For a narrow $[0,0,1]$ quantum well, it reduces to the 2D,  linear momentum dependent 
term ${\cal H}_D={\beta}\left(p_x\sigma_x-p_y\sigma_y\right)/{\hbar}$.
Here, the $\sigma$'s are the Pauli matrices, and $\beta$ is the intensity of this interaction 
(cf Ref.\onlinecite{Ra}). The electric field caused by the structure inversion 
asymmetry (SIA) of the heterostructure generates the  Bychkov-Rashba 
term,  ${\cal H}_R$ \cite{rash}. Since in the asymmetric quantum wells  the SIA 
comes from the vertical direction, the Bychkov-Rashba interaction 
${\cal H}_R$ has the form: 
${\cal H}_R={\alpha}\left(p_y\sigma_x-p_x\sigma_y\right)/{\hbar}$, where 
$\alpha$ is the corresponding strength \cite{Ra}.

The correlation between different spin-dependent 
interactions is a key point in the understanding of spin phenomena in 
semiconductors.
In this work, we shall study a  two-dimensional (2D) semiconductor system
in a magnetic field, described by the Hamiltonian (\ref{ham}), paying special 
attention to the interplay between the Zeeman  and spin-orbit interactions. 
We will show that depending on the orientation of 
the magnetic field and the strength of the spin-orbit coupling
this interplay produces a dramatic effect on the electronic spin structure of 
a 2D heterostructure.

We first consider the case of $B|| [0,0,1]$, i.e., of the
magnetic field parallel to the growth direction (z) of most heterostructures.
In this case the momentum $\vec p$ is replaced by a generalized momentum
$\vec{P}=\vec{p}+(e/c)\vec{A}$,  where the vector potential $\vec{A}$ 
of the vertical magnetic field ${\vec B}=(0,0,B)$  is taken in the symmetric 
gauge $\vec{A}=B/2(-y,x,0)$. In the latter case the Zeeman term is
given by: ${\cal H}_Z=g^*\mu_BB\sigma_z/2={\varepsilon_z}\sigma_z/2$, 
where $g^*$ represents the bulk effective gyromagnetic factor, that is assumed 
to be constant, $\mu_B=e\hbar/2m_ec$ is the Bohr's magneton and $B$ the 
intensity of the applied field.

When a single source of spin-orbit coupling is considered, 
the Rashba or the Dresselhaus one, it can be completely solved \cite{schlie2,das,falko}. 
The Hamiltonian commutes with the total angular momentum operator ($J_z=L_z+S_z$)
when the Bychkov-Rashba interaction is only considered. On contrary, when 
the Dresselhaus term presents alone, the symmetry operator is given by 
$L_z-S_z$ \cite{valin}. If both terms are considered, no symmetry is available 
and a solution is unknown. Below we propose an approximate solution for the 
full problem, which is one of the main results of the paper.

In the absence of any spin-orbit coupling, one obtains Landau levels spaced in 
energy by the cyclotron frequency ($\omega_c=eB/m^*c$) 
and having a spin splitting given by the Zeeman interaction. To include the effects
of the spin-orbit coupling, we transform the Hamiltonian Eq.(\ref{ham}) 
to the reference frame where it is diagonal in a spin space up to a second order
\begin{equation}
\label{tra}
{\cal H}^{\prime}={\cal U}^{\dagger}{\cal H}{\cal U},\:\:{\cal U}=e^{-i(\gamma\hat{O}+\delta\hat{Q})}.
\end{equation}
The explicit expressions of the operators involved in the unitary transformation read,
\begin{equation}
\label{op}
\hat{O}=P_x\sigma_x+P_y\sigma_y\,,\,  \hat{Q}=P_y\sigma_x+P_x\sigma_y,
\end{equation}
and
\begin{equation}
\label{par}
\gamma=-\frac{\alpha/\hbar}{\hbar\omega_c-\varepsilon_z},\: 
\delta=\frac{\beta/\hbar}{\hbar\omega_c+\varepsilon_z}.
\end{equation}

We expand the transformed Hamiltonian, Eq.(\ref{tra}), up to the second order in 
the parameters of the transformation. Third order terms can be neglected if the 
magnitude of the operator involved in the transformation is much smaller than unity. 
This requirement is satisfied for the spin-dependent terms that are dominated by 
the orbital effect of the magnetic field. Therefore, the validity of the present 
approach is roughly restricted to the regime where the energy scale concerning
the spin-dependent interactions ($\varepsilon_z,\alpha,\beta$) is much
smaller than the energy scale given by the cyclotron frequency ($\hbar\omega_c$). 
Nevertheless, note that no restriction is imposed on the relative weight
of the spin-orbit coupling and the Zeeman interaction.

Under the above hypothesis, the resulting transformed Hamiltonian is diagonal in 
spin space 
\begin{eqnarray}
\label{htra}
{\cal H}^\prime&=&\frac{1}{2m^*}\left[1-\frac{m^*}{\hbar^2}
                  \left(\frac{\alpha^2}{\hbar\omega_c-\varepsilon_z}-\frac{\beta^2}{\hbar\omega_c
		  +\varepsilon_z}\right)\sigma_z \right](P_x^2+P_y^2) \nonumber \\
		&&+\frac{\varepsilon_z}{2}\sigma_z -
		\frac{m^*}{\hbar^2}\left(\frac{\alpha^2}{1-\varepsilon_z/\hbar\omega_c}
		-\frac{\beta^2}{1+\varepsilon_z/\hbar\omega_c} \right) \nonumber \\
		&&+2\frac{\alpha\beta}{\hbar^2}\frac{\varepsilon_z}{\varepsilon_z^2-
		(\hbar\omega_c)^2}(P_xP_y+P_yP_x)\sigma_z
\end{eqnarray}

 and the corresponding eigenstates have their spin aligned in the vertical 
 direction ($\sigma_z$). The last term in Eq.(\ref{htra}) represents the  
 contribution of the interplay between the whole set of spin-dependent 
 interactions: Dresselhaus, Bychkov-Rashba and Zeeman. It is diagonal in spin, 
 although it has a crossed dependence on the in-plane components of
 the generalized momentum. To illuminate the relevance of this term 
 to the spectrum, we apply a second transformation to the 
 Hamiltonian, Eq.(\ref{htra}),
\begin{equation}
\label{stra}
{\cal U}_s=e^{-i\eta(P_y^2-P_x^2)}\,,\,\eta=\frac{\alpha\beta}{\hbar^2}
\frac{\varepsilon_z}{\hbar\omega_c(\varepsilon_z^2-(\hbar\omega_c)^2)}
\end{equation}

and keep all contributions, up to the second order. This approximation
is well justified within our approach, since the magnitude of the term is much 
smaller than the corresponding  orbital effect of the magnetic field. In this way, 
the last term in Eq.(\ref{htra}) is transformed to a kinetic-like one:
$\frac{m^*}{\hbar^4}\left(\frac{2\alpha\beta\varepsilon_z}{(\hbar\omega_c)^2-
\varepsilon_z^2}\right)^2(P_x^2+P_y^2)\sigma_z$.
The intensity of this term has a sixth order dependence on the parameters of 
the spin-dependent interactions ($\alpha,\beta,\varepsilon_z$). It is 
much smaller than the first kinetic terms in Eq.(\ref{htra}). 
As a consequence, this term has no relevant  effect on the spectrum and can 
be safely neglected.

Under this approximation, one can see that the main effect of the spin-orbit coupling 
and its interplay with the Zeeman interaction is contained in the first term of 
the Hamiltonian (\ref{htra}). In the transformed reference frame, this term is 
composed of the initial generalized kinetic term plus a spin-dependent one. 
The third term is just a constant that redefines the origin of the 
energy, whereas the second one corresponds to the Zeeman interaction
associated with the vertical field. 

Comparing with the usual expression for the kinetic energy, one observes 
that the spin-orbit interaction cause a redefinition of the effective mass of 
the system. This effective mass ($m_{\uparrow,\downarrow}^*$) depends now on 
the spin orientation of the electrons as well as on the parameters characterising 
the different spin-dependent interactions. In this way we face with a Landau problem
of free electron motion (with this effective mass) in the magnetic field. 

The spectrum associated with the Hamiltonian, Eq.(\ref{htra}) without the last term, 
can be obtained straightforwardly and it reads
\begin{eqnarray}
\label{spec}
\varepsilon_{nls}&=&(2n+\left|l\right|+l+1)\hbar\omega_{c,s}
+\frac{\varepsilon_z }2s\nonumber\\
&-&\frac{m^*}{\hbar^2}\left(\frac{\alpha^2}{1-\varepsilon_z/\hbar\omega_c}-
\frac{\beta^2}{1+\varepsilon_z/\hbar\omega_c}\right),
\end{eqnarray}
where
\begin{equation}
\hbar\omega_{c,s}=\frac{\hbar\omega_c}{2}-\frac{m^*}{\hbar^2}
\left[\frac{\alpha^2}{1-\varepsilon_z/\hbar\omega_c}-
\frac{\beta^2}{1+\varepsilon_z/\hbar\omega_c}\right]s
\label{ncf}
\end{equation}
and $s=\pm 1$ denotes the different spin states.
In Eq.(\ref{spec}), the expression '$(2n+\left|l\right|+l+1)$' is usually replaced by 
'$(2N+1)$' corresponding to the principal Landau level index. The second term is 
the usual Zeeman contribution and the last one represents the constant shift.
Note, that the correction introduced by the spin-orbit coupling in the spectrum,
Eq.(\ref{ncf}), depends quadratically on the intensities of the different 
spin-orbit terms, in contrast with the situation at zero magnetic field, 
where the effect in the  spectrum depends linearly on these intensities. 
For a quantum dot, when only Zeeman and Rashba terms are
considered \cite{qd}, the interplay between the above terms modifies
the spin-orbit intensity, while the kinetic term manifests a weak
dependence on the spit-orbit coupling. 
For $g^*=0$ (Zeeman term is absent) in a quantum dot, the kinetic term and 
the spin-orbit coupling remain without modifications.

When the Zeeman term is absent in the 2D system, the resulting spectrum 
shows, however, an effective spin-dependent cyclotron frequency, 
$\omega_{c,s}=\omega_c-2m^*(\alpha^2-\beta^2)s/\hbar^3$. 
Materials with the effective $g^*=0$ can be created
by changing the concentration of one of the components of a sample 
(see, for example, \cite{salis}).
Comparing the latter 
expression with that given by Eq. (\ref{ncf}), it can be seen that the 
different spin-orbit terms (Bychkov-Rashba and Dresselhaus) 
interact in a different way with the Zeeman one. In both cases, the effect is 
quantified by the ratio between the Zeeman energy and the conventional 
cyclotron frequency (=$\varepsilon_z/\hbar\omega_c$). Note, that this
ratio is independent on the strength of the magnetic field, since it
can be expressed in terms of the 
gyromagnetic factor and the effective mass
\begin{equation}
\frac{\varepsilon_z}{\hbar\omega_c} = \frac{1}{2}g^*\frac{m^*}{m_e}
\end{equation}
Therefore, the interplay acts, in an effective way, renormalising the intensities 
corresponding to the different spin-orbit terms. 
The character of this renormalisation depends
as well as on the sign of the gyromagnetic factor of the semiconductor material. 
If the sign of the gyromagnetic effective factor is negative then the effective 
strength of the Bychkov-Rashba term is reduced, whereas the corresponding to 
the Dresselhaus term is enhanced. On contrary, if the sign of '$g^*$' is positive, 
the effective Bychkov-Rashba term is enhanced and the Dresselhaus one tends to 
be suppressed. Thus, our analytical results provide a transparent explanation for 
numerical results obtained for 2D electron gas with Rashba spin-orbit coupling 
only \cite{usaj} and predict various new phenomena related to the interplay 
between both spin-orbit  and Zeeman terms.

\begin{table}
\begin{ruledtabular}
\begin{tabular}{cccc}
 &$g^*$ &$m^*/m_e$ &$\varepsilon_z/\hbar\omega_c$\\
\hline
GaAs & -0.44 & 0.067 & -0.015\\
InAs & -14.9 & 0.023 & -0.169\\
InSb & -51.6 & 0.014 & -0.355\\
AlAs & 1.52 & 0.15 & 0.112 \\
In$_{0.53}$Ga$_{0.47}$As & -4.38 & 0.038 & -0.082 \\
\end{tabular}
\end{ruledtabular}
\caption{\label{table1} Characteristic values of the parameters involved in the 
ratio between the Zeeman energy and the cyclotron frequency for different 
semiconductor materials.}
\end{table}

In table \ref{table1} there are listed the typical values of the effective gyromagnetic 
factor and the effective mass corresponding to different semiconductor materials. 
A wide gap material such as GaAs, characterized by a relatively weak spin-orbit 
interaction, exhibits a small correction of the spin-orbit effect of about $1.5 \%$. 
However, in the case of narrow gap semiconductors, which are characterized by 
strong spin-orbit intensities and g-factors, the interplay between both interactions
yields a sizeable correction of the effect corresponding to the different 
spin-orbit terms of about $17 \%$ for InAs and $35 \%$ for InSb.

If the magnetic field is parallel to the plane of an electron 
motion, the situation changes substantially. In 2D systems, the orbital effect of 
the applied in-plane magnetic field is frozen due to the strong confinement in the vertical 
direction. The only effect of the field appears through the Zeeman interaction. 
In this case, the Hamiltonian, Eq.(\ref{ham}), preserves the translational 
invariance and the momentum is conserved. The Zeeman term now reads
\begin{equation}
{\cal H}_Z=\frac{\varepsilon_z}{2}\left(\cos\theta\sigma_x+\sin\theta\sigma_y\right),
\end{equation}
where the angle $\theta$ represents the azimuthal orientation of the magnetic field 
in the $x-y$ plane. The corresponding spectrum is given by the 
following analytical expression:
\begin{eqnarray}
\varepsilon_{k_x,k_y,s}=\frac{\hbar^2\vec{k}^2}{2m^*}+s\left[A^2+B^2\right]^{1/2} \nonumber \\
A=\alpha k_y +\beta k_x + \varepsilon_z \cos\theta/2 \nonumber \\
B=\alpha k_x +\beta k_y - \varepsilon_z \sin\theta/2
\end{eqnarray}
where '$k_x$' and '$k_y$' are the components of the in-plane momentum.
This is a contrast to a quantum dot, where the approximate analytical solutions 
are available only for ${\cal H}_Z=0$ and for 
${\cal H_{R,D}}\ll {\cal H}_Z$ \cite{valin}.

To elucidate the implications of the above expression, let us analyse 
some illustrative limit cases. 
First, we will consider the case when a single 
source of spin-orbit coupling is present. In the absence of the magnetic field, 
the spectrum is defined by 
$\varepsilon_k=\hbar^2k^2/2m^*\pm\alpha_i k$ ($\alpha_i=\alpha,\beta$). 
The spin splitting '$\alpha_i k$' is isotropic, i.e., it doesn't depend on the 
orientation of the momentum. However, this symmetry disappears when the Zeeman 
term is included. In the limit $\beta=0$ the spin splitting can be written as
\begin{equation}
\label{sim}
s\left[\left(\alpha k + \frac{\varepsilon_z}{2}\right)^2+
\alpha\varepsilon_z k(\sin(\phi-\theta)-1)\right]^{1/2},
\end{equation}
where the angle '$\phi$' represents the  azimuthal orientation of the momentum in the 
x-y plane. Note that, when the momentum is perpendicular to the magnetic field 
($\phi=\pi/2+\theta$), the spin-orbit coupling and the Zeeman terms are decoupled 
in the spectrum. The same happens for the Dresselhaus interaction, but, in this case, 
when the momentum and the applied field satisfy the condition ($\theta=-\phi$). 
In general, however, the interplay between the Zeeman interaction and the 
spin-orbit coupling is reflected by means of  Eq. (\ref{sim}). 

Let us consider the limit of small Zeeman energy compared to 
'$\alpha k$', which is usually satisfied near the Fermi level of typical 
semiconductor heterostructures. Expanding Eq. (\ref{sim}), we obtain
\begin{equation}
\varepsilon_{k_x,k_y,s}\simeq \frac{\hbar^2k^2}{2m^*}+
s\left(\alpha k +g_{eff}^*(\theta,\phi)\frac{\mu_BB}{2}\right),
\label{pss}
\end{equation}
where 
\begin{equation}
g_{eff}^*(\theta,\phi)=g^*\sin(\phi-\theta)
\end{equation}
is an anisotropic effective gyromagnetic factor 
originated due to the interplay between the two spin-dependent interactions.
Thus, at $B\neq 0$, the splitting partners, determined 
by the term $s\alpha k$ in Eq.(\ref{pss}), are enriched
by the second term that is linear dependent on the magnetic field.
Due to a random orientation of an electron momentum in plane, 
the borders of such sub-bands are restricted within the range 
$\left[-g^*\mu_BB/2, +g^*\mu_BB/2 \right]$, corresponding
to the different momentum orientations '$\phi$', instead of the discrete spin states
$\pm g^*\mu_BB/2$ associated with an isotropic g-factor.
Depending on the relative orientation between momentum and the applied 
field, the value of the effective g-factor can change its sign with respect to 
the initial bulk value.  In fact, in  experiments  with
a tilted magnetic field \cite{Sin} it was observed that the effective g-factor depends
on the tilted angle in the growth direction (z). 
In this case the increase/decrease  of the g-factor
may be influenced  by the orbital motion alone, since the tilting of the
magnetic field affects different components of the orbital momentum.
In contrast to this experiment, we provide a transparent mechanism of 
the anisotropy of the effective g-factor  
{\it in plane} due to the interplay between magnetic field and strong spin-orbit coupling.  
This phenomenon may serve as a guideline to construct a
novel class of read/write devices based on nanostructures with the
strong spin-orbit interaction.

In the opposite limit $\varepsilon_z\gg \alpha_i k$, the spectrum 
can be approximated by
\begin{equation}
\varepsilon_{k_x,k_y,s}\simeq \frac{\hbar^2k^2}{2m^*}+
s\left(\alpha k\sin(\phi-\theta) +\frac{g^*\mu_BB}{2}\right).
\end{equation}
Consequently, in this limit, there is an angular dependence of the orbital 
contribution to the spin splitting, 
which can be attributed to an anisotropic effective strength of the Bychkov-Rashba 
term.

Another interesting limit is that  the intensities of both spin-orbit terms 
are equal ($|\alpha|=|\beta|$).  Assuming $\alpha=-\beta$, the spin splitting 
reads
\begin{equation}
s\left[\left(\sqrt{2}\alpha k_{yx} + \frac{\varepsilon_z}{2}\right)^2+
\sqrt{2}\alpha\varepsilon_z k_{yx}
\left(\cos\left(\theta-\frac{\pi}{4}\right)-1\right)\right]^{1/2},
\end{equation}
where $k_{yx}=k_y-k_x$. When the magnetic field is oriented at $\theta=\pi/4$, 
the spin-orbit and the Zeeman terms don't couple in the spectrum, independently 
on the momentum's orientation. In this case the Hamiltonian has a symmetry 
in a spin space and, consequently, orbital motion and spin are effectively 
decoupled and the spin-orbit interaction is effectively suppressed. 
Thus, we generalize the result found at $B=0$ by Schliemann {\it at el} \cite{schlie}
for a nonzero magnetic field.

In summary, we have analytically studied the effect of an applied magnetic field
on the interplay between the Zeeman, Bychkov-Rashba and Dresselhaus
interactions in a 2D electron system. 
When the magnetic field is perpendicular to the plane, we found that
the spin-orbit interaction induces a renormalization of the effective mass that becomes
dependent on the spin orientation of the electrons. This spin-dependent contribution
is noticeably influenced by the interplay between the Zeeman interaction and the spin-orbit
coupling, which modifies in a different way the effective strength corresponding 
to each spin-orbit term.
The spin-orbit coupling effectively transforms to a spin-dependent cyclotron frequency 
($\omega_{c,s}=eB_z/m_s^* c$) that imposes different spatial scale of an electron motion.
Indeed, the spin separation of cyclotron motion has been observed recently in \cite{rok} 
but in case of 2D {\it hole} gas, which exhibits a stronger spin-orbit interaction.
The magnitude of the interplay between both spin-dependent interactions has a sizeable 
value for a narrow gap semiconductors with large g-factors. 
For the in-plane magnetic field, the interplay induces the 
anisotropic effective gyromagnetic factor that would lead to the observation of a
continuum of Zeeman sub-levels in the spin structure of the spectrum. 
When the Zeeman interaction dominates
over the spin-orbit coupling, an anisotropic effective spin-orbit intensity emerges.

This work was partially supported by Grants PRIB-2004-9765 (Govern de les Illes
Balears) and  BFM2002-03241 (MEC). R. G. N.  acknowledges the support from 
the Ram\'{o}n y Cajal programme (Spain).

\end{document}